\documentclass[twocolumn,aps,pre]{revtex4}
\usepackage{graphicx,color}
\usepackage{pdfpages}
\begin{document}
\title{\bf{Global Density Profile  for  Asymmetric 
Simple Exclusion Process from Renormalization Group Flows}}
\author{Sutapa Mukherji${}^*$ and  Somendra M. Bhattacharjee${}^{**}$}
\affiliation{${}^*$Protein Chemistry and Technology, Central Food
  Technological Research Institute, Mysore 570 020, Karnataka, India\\
${}^{**}$Institute of Physics, Bhubaneswar 751005, India}
\date{\today}
\begin{abstract}
  The totally asymmetric simple exclusion process along with particle
  adsorption and evaporation kinetics is a model of boundary-induced
  nonequilibrium phase transition.  In the continuum limit, the
  average particle density across the system is described by a
  singular differential equation involving multiple scales which lead
  to the formation of boundary layers (BL) or shocks.  A
  renormalization group analysis is developed here by using the
  location and the width of the BL as the renormalization parameters.
  It not only allows us to cure the large distance divergences in the
  perturbative solution for the BL but also generates, from the BL
  solution, an analytical form for the global density profile.  The
  predicted scaling form is checked against numerical solutions for
  finite systems.
\end{abstract}

\maketitle
\section{Introduction}

The totally asymmetric simple exclusion process (TASEP) is a model for
boundary-induced nonequilibrium phase transitions
\cite{Golinelli,liggett,derrida}, though it had its genesis in
modeling polymerization on biopolymeric templates
\cite{gibbs1,gibbs2}.  In this open, driven system, particles,
representing biomolecules, hop in a specific direction on a
one-dimensional lattice, obeying a mutual exclusion rule forbidding
double occupancy of any site.  The rates of injection and withdrawal
of particles at the boundaries are the drives necessary to maintain
the system in a nonequilibrium steady state and they determine the
 bulk  properties, for example, the average particle density in the bulk, 
 in the steady-state.   Unlike equilibrium systems, there is
a bulk-boundary duality and the bulk transitions are completely
encoded in thin boundary layers (BL) of the particle density.  BLs are
not just microscopic details because they survive the continuum limit
which washes out some small-scale details.  This unusual feature of
the steady-state transitions has motivated many studies that involve
developments of new methods \cite{schuetz,popkov,smfixedpt} and new
models \cite{evans,frey,evans1} with an aim to understand
nonequilibrium phase transitions, and to obtain the phase diagram in
the parameter space of the problem.  The primary issue in such
problems is that the resulting differential equation for  the  density
profile is a stiff one involving multiple scales, making it very hard
to solve in the bulk limit with the given boundary conditions.  We
show how the stiffness of the equation can be harnessed with the help
of renormalization group (RG) to develop an interpolation scheme for
the density profile from finite size to the bulk limit.

The usual procedure of the boundary layer analysis\cite{smsmb,cole}
for problems with multiple scales involves asymptotic matching of
different parts of the solutions obtained for different scales.  More
specifically, the rapidly varying BL solution and the smoothly varying
bulk solution must match in the overlapping asymptotic limits
\cite{cole}.  Such an approach, in an order-by-order scheme for
separate scales, ultimately leads to nonphysical divergences, which
need to be handled by an RG analysis.  The power of the RG approach as
a tool for asymptotic analysis has been illustrated in Refs.
\cite{oono} and \cite{mallet} for different types of nonlinear
problems.  We develop a procedure where the width of the BL is taken
as the parameter to be renormalized to remove the divergences with the
help of an arbitrary length scale $\mu$ that adjusts the location of
the BL.  The condition that the density profile should not depend on
$\mu$, then yields the RG equation for the width.  The solution of the
RG equation allows us to reconstruct the density profile.
  
\section{TASEP with nonconservation kinetics}
Let us consider TASEP with an additional adsorption and evaporation of
particles to and from the lattice (Langmuir kinetics (LK)).  The
dynamics of particles can be represented by a master equation that
describes the time evolution of the occupancy variable $\tau_i$ taking
values $1$ or $0$ depending on whether the $i$th site is occupied or
empty, respectively. The master equation is
\begin{eqnarray}
\frac{d\tau_i}{dt}=\tau_{i-1}(1-\tau_i)-\tau_i(1-\tau_{i+1}) +
\omega_a(1-\tau_i)-\omega_d \tau_i, \label{master}
\end{eqnarray} 
where the first two terms on the right hand side of Eq.~(\ref{master})
represent the hopping of particles to the empty forward 
site, and the last two terms represent adsorption and evaporation of
particles with rates $\omega_a$ and $\omega_d$, respectively
\cite{adsorp}.  For a finite lattice of $N$ sites, particles are
injected at $i=1$ at rate $\alpha$ and are withdrawn from the lattice
at $i=N$ at rate $\beta$.  The time evolution of the average particle
density, $\langle \tau_i\rangle$ at a given site $i$ can be found from
the statistical averaging (denoted by $\langle ..\rangle$ ) of the
above equation.  In a mean-field approximation, factorizing the
correlations such as $\langle\tau_i\tau_j\rangle \approx
\langle\tau_i\rangle\langle\tau_j\rangle$, 
the steady-state density 
in the continuum limit (the lattice spacing, $a\rightarrow 0$, and
$N\rightarrow \infty$ with $Na=1$), is described by
 \begin{eqnarray}
 \epsilon\frac{d^2\rho}{dx^2}+(2\rho-1) \frac{d\rho}{d x}+
\Omega  (r-(r+1)\rho)=0, \label{steadyrho1}
\end{eqnarray}
where $x$ denotes the location along the lattice,
$\rho(x)=\langle\tau_i\rangle$ is the average density, and $\epsilon$
is a small parameter proportional to the lattice spacing. The boundary
conditions (BC) are $\rho(x=0)=\alpha,$
{\rm and} $\rho(x=1)=1-\beta=\gamma.$ To obtain Eq.
(\ref{steadyrho1}), the neighboring densities are written as
$\rho(x\pm a)=\rho(x)\pm
a\frac{d\rho}{dx}+\frac{a^2}{2}\frac{d^2\rho}{dx^2} ...$, and we took
$\Omega_a=\omega_aN$, $\Omega=\Omega_d=\omega_dN$ and
$\Omega_a/\Omega_d=r$. This limit of $\Omega_{a,d}\sim O(1)$ is
required to make the net flow into the system comparable to the
current along the lattice.

The differential equation Eq. (\ref{steadyrho1}) is singular due to
the small prefactor $\epsilon$ in front of the highest order
derivative.  In the extreme limit, $\epsilon=0$, the equation reduces
to a first order equation which cannot, in general, satisfy two
BCs.  The loss of one BC               leads to the
appearance of a boundary layer. Another way of seeing this is to
realize that for small but finite $\epsilon$, there are two scales,
$x$, and $\tilde x=x/\epsilon$, so that the density is a function of
two widely different scales, making the equation stiff to solve.
Standard numerical procedures with  special continuation
strategies\cite{cash} fail to 
converge for small $\epsilon$.  To overcome this problem, the
steady-state behaviour of this system has been studied using various
methods such as domain wall theory, boundary layer analysis, numerical
simulations, etc.\cite{popkov,smfixedpt,krug}.

\begin{figure}[htbp]
\includegraphics{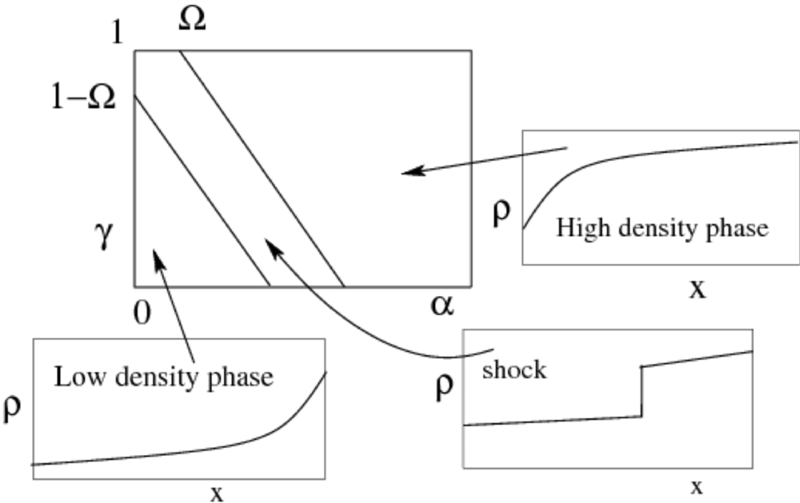}  
\caption{A part of the phase diagram ($\gamma$-$\alpha$) of TASEP with
  Langmuir kinetics with $\omega_a=\omega_d ~(r=1).$ The low-density
  phase to the shock phase boundary is $\gamma=1-\Omega-\alpha$, while
  that for the shock to the high density phase is
  $\gamma=1+\Omega-\alpha$. The density profiles in the three phases
  are shown schematically.  }
\label{fig:phasediag}
\end{figure}

The particle conserving TASEP $(\omega_a=\omega_d=0)$ can exist
broadly in three phases which are low-density, high-density, and
maximal current phases.  With LK, there is a difference in phase
diagrams for $r=1$ and $r\neq 1$. However, for both the cases, there
is a region in the phase diagram where the high density (HD) and
the low-density (LD) phases are separated by a shock phase.  This
region is of interest, see Fig. \ref{fig:phasediag}, and corresponds
to $\gamma>0.5+\Omega$.  For $r=1$, the average particle density 
in the bulk changes linearly with $x$. In  the LD phase for
$\alpha<\frac{1}{2}$, the average density across the lattice remains
less than $1/2$, consistent with the BC at $x=0$, while a BL near
$x=1$ matches the BC at that end.  Such a phase appears for
$\alpha+\gamma<1-\Omega.$ Similarly, for $\beta<1/2$ and
$\alpha+\gamma>1+\Omega$, the system is in an HD phase in which the
major part of the density $>1/2$, consistent with $\gamma>\frac{1}{2},$
with the BL around $x=0$.  The picture remains more or less similar
for $r\neq 1$ in this region, though the bulk density is no longer
linear in $x$.

\section{Density profile via Renormalization Group}
The RG analysis is based on the boundary layer 
part  of the  particle density profile, and the
outcome is a globally valid solution for the entire density
 profile, thereby
broadening the region of validity of the boundary layer solution. 
We rewrite  Eq.  (\ref{steadyrho1}) as
\begin{eqnarray}
  \frac{d^2\phi}{d\tilde x^2}+\phi \frac{d\phi}{d \tilde x}+
  \epsilon\Omega' (r_0-\phi)=0,\label{eq:1} 
\end{eqnarray} 
where, $\tilde x={x}/{\epsilon},$
\begin{equation}
  \label{eq:3}
\phi=2\rho-1, r_0=\frac{r-1}{r+1}, {\rm and}\,\,  \Omega'=(r+1)\Omega.  
\end{equation}
It is to be noted that Eq. (\ref{eq:1}) remains invariant under a
shift of origin $\tilde x\to \tilde x-x_0$; this symmetry is exploited
below.  We look for a regular perturbative solution of the form,
\begin{equation}
  \label{eq:4}
\phi=\phi_{0}+\epsilon\phi_{1}+... \,.   
\end{equation}
The zeroth order solution is 
\begin{equation}
  \label{eq:2}
\phi_0(\tilde{x})=p\tanh\left[\frac{p}{2}(\tilde x+k)\right], \label{phi00}
\end{equation}
which is characterized by two parameters $k$ and $p$, related to the
centre and the width of the boundary layer, respectively.  In the
boundary layer approach, this $\phi_0$ is the BL on the scale of
$\tilde x$, to be matched with the bulk solution of
Eq.~(\ref{steadyrho1}) for $\epsilon=0$.  We, instead, extend the 
BL solution to the  next order.  At $O(\epsilon)$ level, $\phi_1$ satisfies
the equation
\begin{eqnarray}
\frac{d^2\phi_1}{d\tilde x^2}+\frac{d(\phi_0\phi_1)}{d\tilde x}+\Omega'(r_0-\phi_0)=0.\label{phi11}
\end{eqnarray}
The divergence mentioned earlier can now be seen from
Eq. (\ref{phi11}).  It shows that $\phi_1\sim \tilde x$, for $\tilde
x\to\infty$, due to the limit $\phi_0\to p$.  To identify this
divergence in $\phi$, let us redefine
$$\tilde x=\frac{x-{x}_0}{\epsilon}, {x}_0=\epsilon k, 
\,\,{\rm and \,\,}{\psi}(p,\tilde
x)=\tanh\left(\frac{p}{2}\tilde x\right).$$ 
The solution, $\phi$, upto $O(\epsilon)$, is given by
\begin{eqnarray} 
  &&\phi=p \;{\psi}(p,\tilde x)-\epsilon \frac{r_0-p}{p} 
\Omega' \;\tilde x\; \psi(p,\tilde x)+ \epsilon {\cal R}, \label{fullsoln}
\end{eqnarray} 
where only the diverging terms at the $O(\epsilon)$ level are shown
explicitly with $\epsilon{\cal R}$ representing all the regular terms
together.  
See Appendix A and B for details.  
This naive
perturbation theory shows inconsistency as $\tilde x\rightarrow
\infty$ since in this limit, the term at $O(\epsilon)$ level in Eq.
(\ref{fullsoln}) becomes comparable to the zeroth order term.  The
divergence appearing in Eq. (\ref{fullsoln}) can be isolated by
introducing an arbitrary length scale $\mu$ that adjusts the location
of the BL, to write Eq. (\ref{fullsoln}) as
\begin{eqnarray}
\phi&=&p\psi(p,\tilde x)-\epsilon \frac{r_0-p}{p} 
\Omega' (\tilde x-\mu) \psi(p,\tilde x) -\nonumber\\
  &&\quad\epsilon\mu\frac{r_0-p}{p}\Omega'\psi(p,\tilde x)+\epsilon
  {\cal R}. \label{fullsoln1} 
\end{eqnarray} 
In the next step, we introduce a renormalized parameter $p_r$ defined
through the equation $p=p_r(\mu)+\epsilon a_1(\mu)$ to absorb the 
divergence in Eq. (\ref{fullsoln1}).  Therefore, we set 
\begin{eqnarray}
a_1=\frac{\mu}{p_r}(r_0-p_r)\Omega'.
\end{eqnarray} 
The divergence-free density profile now appears as 
\begin{eqnarray}
  \phi&=&p_r\psi(p_r,\tilde x)+\epsilon \frac{\mu}{2}(r_0-p_r)
 {\rm sech}^2[\frac{p_r\tilde  x }{2}]
\Omega' \tilde x-
\nonumber\\
  &&   \frac{\epsilon}{p_r}\psi(p_r,\tilde x)\; (r_0-p_r)\Omega'\;(\tilde x-\mu),\label{fullsoln2}
\end{eqnarray}
where the sech term comes from the Taylor expansion of $\psi$.
Since the final density profile must be independent of the arbitrary
length scale $\mu$, we must have ${\partial \phi}/{\partial
  \mu}=0$. The complete expression of ${\partial \phi}/{\partial
  \mu}$ along with cancellations necessary to ensure that
${\partial \phi}/{\partial \mu}$ is zero at $O(\epsilon)$ level is
shown in Appendix B. 
This condition leads to the
renormalisation group equation to $O(\epsilon)$ as
\begin{eqnarray}
  \frac{dp_r}{d\mu}=-\epsilon\; \frac{r_0-p_r}{p_r} \;\Omega'.\label{rgeqn1}
 \end{eqnarray}
It is interesting to note that this RG equation is the bulk equation,
Eq. (\ref{steadyrho1}) with $\epsilon=0$, when expressed in terms of
$\phi$, and $\mu$ replacing $x$.   
   
For $r=1$ (i.e.,  $r_0=0$),   the solution of
Eq.~(\ref{rgeqn1}) is  $p_r=2\Omega\epsilon\mu+c$, $c$ being a constant.  
Substituting this in (\ref{fullsoln2}) along with $\mu=\tilde x$, we have 
the density profile, to  leading order, as
\begin{eqnarray}
&& \phi(x)=(C+2\Omega x)\; \tanh[(C+2\Omega x)\; {\tilde x}/{2}], \label{eqmain:10} 
\end{eqnarray}
where $C$ is a constant to be determined, and $x_0$ in  $\tilde
x=(x-x_0)/\epsilon$ as the other unknown constant.  These two
constants are determined by the boundary conditions.

In case of $r\neq 1$, we  solve Eq. (\ref{rgeqn1}) perturbatively
for small $r_0$.  Expressing Eq. (\ref{rgeqn1}) in terms of
$\lambda= \epsilon\Omega' r_0$, we obtain a perturbative solution for
$p_r$ with $p_r=p_r^0+\lambda p_r^1+O(\lambda^2)$ as
\begin{eqnarray}
p_r=\Omega'\epsilon \mu+c-r_0 \ln(\Omega'\epsilon\mu+c)+O(\lambda^2),\label{eqmain:11}
\end{eqnarray}
where $c$ is a constant.  Replacing $\mu$ by $\tilde x$, we have the
final form of the density profile as  
\begin{eqnarray}
\phi&=&\left[\Omega' x+C-r_0 \ln(\Omega'  x +C)\right] \times\nonumber\\
&&\tanh\left[\{\Omega'  x+C-r_0 
\ln(\Omega' x+ C)\}\; \tilde x/2\right]. \label{eqmain:12}
\end{eqnarray}

Eqs. (\ref{eqmain:10}) and (\ref{eqmain:12}) are the main results of
our paper.  The bulk solutions can be found from these equations by
considering $\tilde x\rightarrow \infty$ limit in which
$\tanh\rightarrow 1$. In case of $r=1$, $\phi$ approaches a linear
function of $x$ as obtained from the boundary layer analysis in Ref.
\cite{evans}.  In case of $r\neq 1$, the density profile $\phi$ in the
$\tilde x \rightarrow \infty$ limit recovers the bulk solution which
has a nonlinear dependence on $x$.  The boundary layer parts, on the
other hand, can be found from the $\tilde x\rightarrow 0$ limit of
expressions in Eqs. (\ref{eqmain:10}) and (\ref{eqmain:12}). As
Eq.~(\ref{eqmain:10}) shows, in the leading order, the density profile
has a form $C\tanh{\big(}C \tilde x/2{\big)}$ in agreement with the
results obtained through the boundary layer analysis \cite{smsmb}.
Interestingly, the RG analysis, via the renormalization of the width
because of adsorption/desorption kinetics of particles, leads to
further subleading correction terms which contribute for finite
$\epsilon$.  Instead of a simple additive form for the density over
two scales, we see a more complex solution where the local bulk
density affects the ``local'' width of the  boundary
layer.  The $\epsilon$-dependent term in Eq. (\ref{steadyrho1}) comes
from the diffusive contribution to the current, and therefore it is
significant only in the region of rapid variation as in a BL or a
shock. To leading order in the boundary layer analysis, this thin
region does not generate much current from the Langmuir kinetics.  As
the BL thickens for not-so-small $\epsilon$, there is an appreciable
contribution from the $\Omega$-dependent kinetics.  Our RG analysis
captures this aspect of the problem.  Herein lies the importance of
Eqs.  (\ref{eqmain:10}) and (\ref{eqmain:12}), which provide an
interpolation formula from finite $\epsilon$ to the bulk.

\begin{figure}[htbp]
\includegraphics{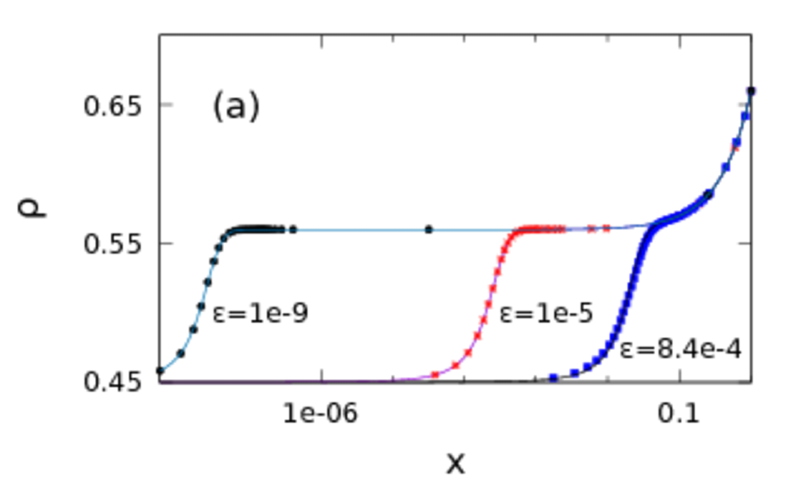}  
\includegraphics{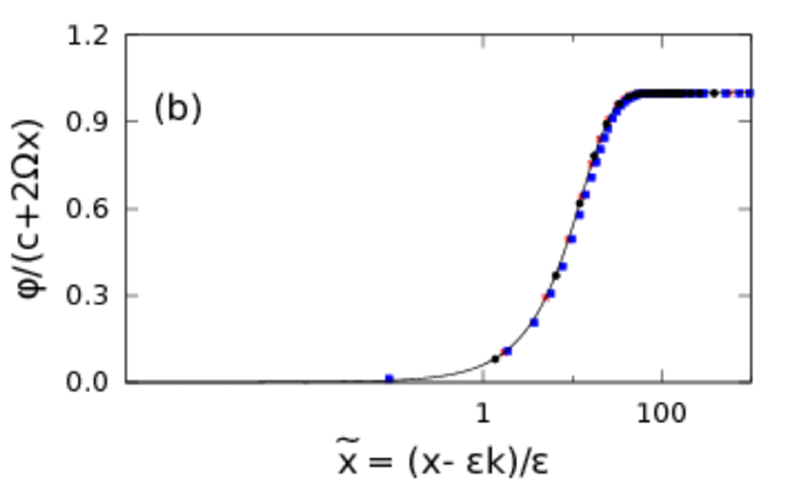}  
\caption{RG solution vs numerical results. (a)Results for
  $\epsilon=8.4\times 10^{-4}, 10^{-5}, 10^{-9}$.  Data points are
  from numerical solutions of Eq. (\ref{steadyrho1}) for $r=1,
  \Omega=0.1, \alpha=0.45$ and $\gamma=0.66$.  The RG solution of Eq.
  (\ref{eqmain:10}) are shown by solid lines.  (b) Data collapse plot
  for the same set of data as in (a) (same symbols).  Labels show the
  variables along x and y axes. The solid line is the tanh part of Eq.
  (\ref{eqmain:10}).}\label{fig:2}
\end{figure}

\subsection{Comparison with numerics}

We compare the numerical solution of Eq.~(\ref{steadyrho1}) for $r=1$,
with plots obtained from the RG solution, Eq. (\ref{eqmain:10}). In
Fig. \ref{fig:2}, plots for the high-density phase with the boundary
conditions $\alpha=0.45$ and $\gamma=0.66$ are shown.  The numerical
solutions of the full differential equation for three different
$\epsilon$, viz., $\epsilon=8.4\times 10^{-4},10^{-5}, 10^{-9}$, are
shown here. For the RG solution in Eq. (\ref{eqmain:10}), the constants
$C$ and $x_0=\epsilon k$ are found from the boundary conditions at
$x=1$ and $x=0$.  This is based on the observation that the boundary
layer in the high density phase is formed at the $x=0$ boundary.  The
equations are $C+2\Omega =0.32$ and $C \tanh[C x_0/2\epsilon]=0.1$,
yielding $C=0.12$ and $k=4.98$.  A nice agreement with the RG solution
is seen for the corresponding $\epsilon$'s.  Moreover, 
Eq.~(\ref{eqmain:10})
admits a scaling form via a collapse of all curves for
different $\epsilon$'s if $\phi(x)/(C+2\Omega x)$ is taken as a
function of $\tilde{x}=(x-x_0)/\epsilon$. Such a form is not expected
from the naive boundary layer solution.  A data collapse plot for all
the numerical solutions is shown in Fig. \ref{fig:2}b,  confirming the
predicted scaling.

\begin{figure}
\includegraphics{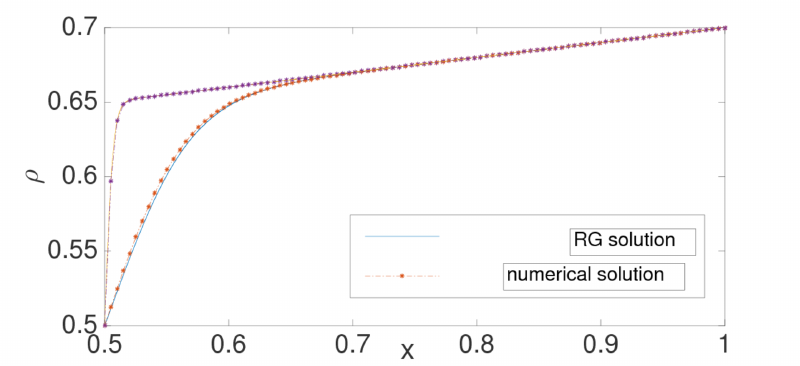}  
\caption{Density profile for shocks. The upper and lower curves
  correspond to $\epsilon=0.001$, and $0.01$, respectively. The
  boundary parameters are $\alpha=0.3$ and $\gamma=0.7$. The graph is
  plotted over the range $x\in [0.5:1]$. }\label{fig:3}
\end{figure}

We also compared the profiles for the shock phase.  For a shock phase
with $x_0$ somewhere in the interior of the lattice, there is a
symmetry $\phi(\tilde x)=-\phi(-\tilde x)$, obeyed by Eq.
(\ref{eq:1}).  We, therefore, concentrate on $\tilde x>0$ only.  The
boundary conditions chosen here are $\alpha=0.3$ and $\gamma=0.7$, so
that the shock is formed at $x_0=0.5$.  Eq. (\ref{steadyrho1}) is
solved numerically with these boundary conditions. With $x_0=0.5$ and
the BC at $x=1$,  we have  $C+2\Omega(x-x_0)=0.4$, so that $C=0.3$.  The symmetry
automatically fixes the boundary condition at $x=0$.  In this way, the
RG analysis performed 
with boundary layer located near one of the boundaries can be utilized
here.  A good agreement is noted between the  numerical solution of the full
differential equation and the RG solution as given in Eq.
(\ref{eqmain:10})  (see Fig. \ref{fig:3}).

\section{Summary}

In this paper, we developed a renormalization group scheme to
determine the particle density profile in a  one-dimensional,
particle non-conserving totally asymmetric simple exclusion process.  The
 particle adsorption/desorption kinetics (Langmuir kinetics)
is the source of  particle non-conservation while the steady state of nonzero current is
maintained by the injection and the withdrawal rates at the
boundaries.  The continuum differential equation for the proocess 
is singular due to a small prefactor ($\epsilon$) in front of its
highest order derivative term that comes from diffusion.  As a
consequence of the singularity, the perturbative solution on the
scale of $\tilde{x}=x/\epsilon$ shows divergences at $O(\epsilon)$ at large
distances, $\tilde{x}\to\infty$.  Upon absorbing the divergences
systematically through 
renormalizations of the position and the
width of the boundary layer, we arrive at a globally valid
density-profile which describes both the boundary layer and its
crossing over to the bulk solution.  One of the predictions of the
solution is the appearance of a finite-size scaling form for the
density, which compares well with the results from 
direct numerical solutions of the steady-state differential equation
in the high-density and the shock phases. We believe our procedure is
general enough to apply to other boundary induced transitions as well.

\begin{acknowledgments}
SM acknowledges financial support from Department of Science and Technology, India through grant number EMR/2016/06266. 
\end{acknowledgments}
\vfill
\appendix


\section{Boundary layer analysis}
The differential equation describing the density is given by 
 \begin{eqnarray}
 \epsilon\frac{d^2\rho}{dx^2}+(2\rho-1) \frac{d\rho}{d x}+
\Omega  (r-(r+1)\rho)=0. \label{eq:5}
\end{eqnarray}
It is singular due to the small prefactor $\epsilon$ in front of the
highest order derivative.  In the extreme limit, $\epsilon=0$, the
equation reduces to a first order equation which cannot, in general,
satisfy two boundary conditions.  The loss of a boundary condition
leads to the appearance of a boundary layer.  The solution of
(\ref{eq:5}) for $\epsilon=0$ describes the major (bulk) part of
the density profile and in the language of the boundary layer theory,
it is referred as the outer solution.  Thus, at zeroth order in
$\epsilon$, the possible outer solutions $\rho_{\rm out}$ are
\begin{eqnarray}
&& \rho_{{\rm out}}=\Omega x+c,\ \ {\rm and} \ \rho_{{\rm
    out}}=\frac{1}{2}, \  {\rm for}\  \  r=1,  \label{outersol1} \\
\end{eqnarray}
while for $r\neq 1$,
\begin{eqnarray}
&& g(\rho_{\rm out})=\Omega x+c \  {\rm with} \label{transcout}\\
&& g(\rho)=\frac{1}{1+r}{\Big(}2\rho+r_0\log[r-(r+1)\rho]{\Big)}.\label{transc}
\end{eqnarray}
Here $r_0=\frac{r-1}{r+1}$ and $c$ is the integration constant the
value of which can be determined using the boundary conditions.  Equation
(\ref{transcout}) is a transcendental equation which can be solved for
$\rho$ numerically.  An explicit perturbative solution for $\phi$($=2\rho-1$) in
small $r_0$ appears as
 \begin{eqnarray}
 \phi=(\Omega' x+c_0)-r_0 \log(\Omega' x+c_0), 
 \end{eqnarray} 
 where $\Omega'=\Omega(r+1)$ and $c_0$ is a constant.  A boundary layer solution, necessary to
 account for the other boundary condition, can be found out by
 introducing a scaled variable $\tilde x=\frac{(x-x_0)}{\epsilon}$,
 where $x_0$ represents the location of the boundary layer. For
 example, for a boundary layer that appears near $x=1$ boundary,
 $x_0\approx 1$.  With this change of the independent variable,
 equation (\ref{eq:5}), in terms of $\phi$, appears as
\begin{eqnarray}
 \frac{d^2\phi}{d\tilde x^2}+\phi \frac{d\phi}{d \tilde x}+
\epsilon \Omega'  (r_0-\phi)=0.\label{eq:s1}
\end{eqnarray}
In the following, we refer to the solution of
this equation as the inner solution ($\phi_{\rm in}$) or  boundary layer solution.  Upon
one integration, the boundary layer equation at the zeroth order in
$\epsilon$ is $ \frac{d\phi_{\rm in}}{d\tilde x}+\frac{\phi_{\rm
    in}^2}{2}=c_1=\frac{\phi_b^2}{2}$ for all $r$.  Apart from
satisfying a boundary condition, the boundary layer also has to
saturate to the outer solution.  For satisfying two such conditions, a
second order differential equation as (\ref{eq:s1}) is necessary for
the description of the boundary layer.  The second equality in the
differential equation for $\phi_{\rm in}$ ensures that the boundary
layer solution saturates to the outer  solution $\phi_b=2\rho_b-1$ with
$\rho_b$ being the bulk density near the saturation limit. In terms of
$\phi_b$, the zeroth order inner (boundary layer) solution is
\begin{eqnarray}
\phi_{\rm in}=\phi_b \coth[\frac{\phi_b}{2}(\tilde x+k)], \ {\rm or}\ 
\phi_b \tanh[\frac{\phi_b}{2}(\tilde x+k)] \label{tanhbl1}
\end{eqnarray}
where $k$ is the second integration constant.  As can be seen from
(\ref{tanhbl1}), the width of the boundary layer is proportional to
$\epsilon$.  Thus as $\epsilon\rightarrow 0$, a boundary layer
becomes narrower appearing like a jump discontinuity in the density
profile.  In the low-density phase the density profile has a linear outer
or bulk solution along with a boundary layer near $x=1$ ($x_0\approx
1$).  Since the outer  solution, in this case, 
satisfies the boundary condition at $x=0$, we have  $c=\alpha$.
The boundary layer, in this case, satisfies the boundary condition at
$x=1$ and merges to the bulk profile for $x<1$ i.e for $\tilde
x\rightarrow -\infty$. The saturation of the boundary layer and the
bulk solution is ensured through the matching condition
$\rho_{in}(\tilde x\rightarrow -\infty)=\rho_{\rm out}(x=1)=\rho_b$.
For a boundary layer located near $x=0$ boundary (i.e. $x_0\approx
0$), the boundary layer saturates to the bulk in the  $\tilde x\rightarrow
\infty$ limit and satisfies the boundary condition at $x=0$.  The
boundary condition at $x=1$, on the other hand, is satisfied by the
outer solution. Density profiles of this shape are seen in the high-density phase
which is related to the low-density  phase through the particle-hole symmetry.
A more drastic behaviour is seen in the shock phase where the boundary
layer enters into the interior of the system upon deconfinement from
the system boundary.  Such a boundary layer joins a low-density type 
density profile on one side to the high-density type profile on the
other side.

\widetext

\section{Details of the RG calculations}

At the $O(\epsilon^0)$ level, we choose the $\tanh$ type solution for
$\phi_0$.  With this, the complete solution for $\phi_1$ is
\begin{eqnarray}
&& \phi_1(\tilde x)=c_2\  {\rm sech}^2[\frac{p}{2}(k+\tilde x)]+
\frac{1}{4p^2} {\rm sech}^2[\frac{p}{2}(k+\tilde x)] {\Bigg[}2 r_0 \Omega'\cosh[p(k+\tilde x)]+
4\Omega' p\  {\rm polylog}[2,-\exp\{-p(k+\tilde x)\}]-\nonumber\\ && p^2 (k+\tilde x){\Big(}-2\Omega'- r_0 k \Omega'+k \Omega' p+
r_0\Omega' \tilde x+\Omega' p \tilde x-2c_1+  4\Omega' \log[1+\exp\{-p(k+\tilde x)\}]-
4\Omega' \log[\cosh\{\frac{p}{2}(k+\tilde x)\}]{\Big)}-\nonumber\\ && 
2p{\Big (}\Omega'+r_0 \Omega' \tilde x-c_1-
 2\Omega' \log[\cosh\{\frac{1}{2}p(k+\tilde x)\}]{\Big)} \sinh[p(k+\tilde x)]{\Bigg]}.\label{phi1}
\end{eqnarray}
The terms proportional to $\tilde x$ and the $\log[\cosh[...]]$ term
in the last part of the above expression are responsible for the
breakdown of the perturbation series in the $\epsilon\tilde
x\rightarrow O(1) $ limit.  Writing explicitly the diverging terms, one may
write
\begin{eqnarray} 
  &&\phi=p\tanh[\frac{p}{2}(k+\tilde x)]+\epsilon {\Bigg[}-\frac{1}{p}
  \tanh[\frac{p}{2}(k+\tilde x)]
  {\Big(}\Omega'+
  r_0 \Omega' \tilde x-c_1-2\Omega' \log[\cosh\{\frac{p}{2}(k+\tilde x)\}]{\Big)}{\Bigg]}+\epsilon {\cal R},
\end{eqnarray} 
where ${\cal R}$ represents all the regular terms.  Considering the
large $\tilde x$ limit, we rewrite the above expression as
\begin{eqnarray}
  &&\phi=p\tanh[\frac{p}{2}(k+\tilde x)]+\epsilon {\Bigg[}-\frac{1}{p} \tanh[\frac{p}{2}(k+\tilde x)]{\Big(}
  r_0 \Omega' \tilde x-\Omega' p(k+\tilde x){\Big)}{\Bigg]}+\epsilon {\cal R}.
\end{eqnarray} 
The divergences in the last two terms proportional to $\tilde x$ can
be isolated by introducing  an  arbitrary length scale $\mu$.
The renormalized parameter is defined through $p_r$ as
$p=p_r+a_1\epsilon$ with $a_1=\frac{\mu}{p_r}(r_0-p_r)\Omega'$.  Upto
$O(\epsilon)$, the divergence free solution becomes
\begin{eqnarray}
   \phi&=&p_r\tanh[\frac{p_r}{2}(k+\tilde x)]+\epsilon p_r {\rm sech}^2[\frac{p_r}{2} (k+\tilde x)]  \frac{\mu}{2p_r}(r_0-p_r)\Omega'-\nonumber\\
  &&\frac{\epsilon}{p_r}\tanh[\frac{p_r}{2}(k+\tilde x)] (r_0-p_r)\Omega'(\tilde x+k-\mu). 
\end{eqnarray}
Next, we impose the condition that $\phi$ must be independent of the
arbitrary length scale $\mu$ i.e.
$\frac{\partial\phi}{\partial\mu}=0$.  Thus, the following expression
\begin{eqnarray}
  && \frac{\partial\phi}{\partial\mu}=p_r'(\mu) T_h + 
p_r S_h  p_r'(\mu)\frac{(k+\tilde x)}{2}\nonumber\\
  && -{\epsilon }{\Big[}2p_r' S_h T_h  \ \frac{(k+\tilde x)^2}{4}
  {\mu}(r_0-p_r)\Omega'- S_h
     \frac{(k+\tilde x)}{2} (r_0-p_r)\Omega'
  +\mu p_r'\Omega' S_h  \frac{(k+\tilde x)}{2}{\Big]}+\nonumber\\
  &&   \epsilon{\Big[}\frac{1}{p_r^2}p_r' T_h\ ((r_0-p_r)\Omega'(k+\tilde x-\mu))-
  \frac{1}{p_r}p_r' S_h \frac{(k+\tilde x)}{2}  \{(r_0-p_r)\Omega'(k+\tilde x-\mu)\}-\nonumber\\ 
  &&   \frac{1}{p_r}  T_h \{-p_r'\Omega'(k+\tilde x-\mu)-(r_0-p_r)\Omega'\}{\Big]}, \label{appenderivative}
  \end{eqnarray}
where
\begin{equation}
  \label{eq:6}
  T_h\equiv \tanh\left[\frac{p_r}{2}(k+\tilde x)\right],~{\rm and}~ S_h\equiv {\rm sech}^2\left[\frac{p_r}{2}(k+\tilde x)\right]
\end{equation}
satisfies the required condition if, to $O(\epsilon)$, we have 
 \begin{eqnarray}
   \frac{dp_r}{d\mu}=-\frac{\epsilon}{p_r} (r_0-p_r)\Omega'.\label{rgeqnappendix}
 \end{eqnarray}
 In (\ref{appenderivative}), $p_r'(\mu)=\frac{dp_r}{d\mu}$.
Eq. (\ref{rgeqnappendix}) has been solved perturbatively in
small $r_0$ in the main text. 


\pagebreak


\begin{thebibliography}{99}
\bibitem{adsorp}  Adsorption (evaporation) at a site  can take place only if
the site is empty (occupied), i.e. if  $\tau_i=0 (\tau_i=1)$.
\end{thebibliography}
\end{document}